\documentclass[a4paper,twocolumn,english,reprint,superscriptaddress,showpacs,showkeys,longbibliography, nofootinbib]{revtex4-1}
\usepackage{graphicx} 
\usepackage[utf8]{inputenc}
\usepackage{hyperref}
\usepackage{amsmath}

\begin{document}
\title{Constructive plaquette compilation for the parity architecture}

\author{Roeland ter Hoeven}
\thanks{These authors contributed equally to this work}
\affiliation{Parity Quantum Computing GmbH, A-6020 Innsbruck, Austria}
\affiliation{Institute for Theoretical Physics, University of Innsbruck, A-6020 Innsbruck, Austria}

\author{Benjamin E.~Niehoff}
\thanks{These authors contributed equally to this work}
\affiliation{Parity Quantum Computing GmbH, A-6020 Innsbruck, Austria}

\author{Sagar Sudhir Kale}
\affiliation{Parity Quantum Computing GmbH, A-6020 Innsbruck, Austria}

\author{Wolfgang Lechner}
\email{wolfgang@parityqc.com,\\wolfgang.lechner@uibk.ac.at}
\affiliation{Parity Quantum Computing GmbH, A-6020 Innsbruck, Austria}
\affiliation{Institute for Theoretical Physics, University of Innsbruck, A-6020 Innsbruck, Austria}

\newcommand{\TODO}[1]{ { \color{blue} \footnotesize (\textsf{todo}) \textsf{\textsl{#1}} } }

\newcommand{\sigmatz}{\tilde{\sigma}_z}
\newcommand{\psit}{\tilde{\psi}}
\newcommand{\bra}[1]{\langle#1\rvert} 
\newcommand{\ket}[1]{\lvert#1\rangle} 
\newcommand{\qprod}[2]{ \langle #1 | #2 \rangle} 
\newcommand{\braopket}[3]{\langle #1 | #2 | #3\rangle} 
\newcommand{\expect}[1]{ \langle #1 \rangle} 
\newcommand{\cC}{\mathcal{B}}

\date{\today}

\begin{abstract}
  Parity compilation is the challenge of laying out the required constraints for
  the parity mapping in a local way.  We present the first constructive compilation
  algorithm for the parity architecture using plaquettes for arbitrary higher-order optimization problems. This enables adiabatic protocols, where the plaquette layout can natively be implemented, as well as fully parallelized digital circuits. The algorithm builds a rectangular layout of plaquettes, where in each layer of the rectangle at least one constraint is added. The core idea is that each constraint, consisting of any qubits on the boundary of the rectangle and some new qubits, can be decomposed into plaquettes with a deterministic procedure using ancillas. We show how to pick a valid set of constraints and how this decomposition works. We further give ways to optimize the ancilla count and show how to implement optimization problems with additional constraints.
\end{abstract}

\maketitle

\section{Introduction}
Compilation is an important process for near-term quantum devices, where qubit numbers, connectivity and quality of operations are limited \cite{Preskill_nisq, ibm_error_mitigation, quantum_compilation}.  The short decoherence time of such devices places further restrictions on the quantum circuits that can be successfully executed. Therefore, it is important to design algorithms that have an efficient mapping to the available hardware \cite{dwave-2000, google_supremacy, heavy_hex_ibm, quantinuum_racetrack, pasqal_300}. Co-design of quantum
hardware and software can be very fruitful, as hardware-specific optimizations can lead to improved algorithms and new algorithms can motivate hardware improvements. 

In quantum optimization, the solution of hard optimization problems is encoded in the ground state of a Hamiltonian, which is then prepared with a quantum algorithm. The main challenge is that this Hamiltonian may require long-range interactions and large circuit depth. The parity architecture~\cite{LHZ, compiler_paper, plaquette_optimizer, modular_qaoa} offers a solution to this
challenge by utilizing existing devices and offering a blueprint for the
development of future devices. The core idea is to map the difficult to implement Hamiltonian to a Hamiltonian that only requires local interactions and constant circuit depth, at the cost of a qubit overhead.  Over the last few years, several approaches have been developed that use plaquette layouts as a
basis. Quantum circuit implementations of plaquette layouts can be
optimized~\cite{plaquette_optimizer} and algorithm performance can be improved
by enforcing some constraints using dynamics~\cite{modular_qaoa}.  In this work,
we show how to constructively generate plaquette layouts in the parity
architecture for higher-order constrained optimization problems.  

There has been good progress in theory and in experiment to realize the required plaquette interactions on quantum hardware. For example, for analog quantum computing there are developments using superconducting parametric oscillators~\cite{nec_coupler}. For digital quantum computing, gates using superconducting qubits~\cite{superconducting_coupler} and neutral atoms~\cite{rydberg_coupler} have been proposed.  

In Ref.~\cite{compiler_paper}, parity compilation for higher-order optimization
problems was introduced, i.e., problems with a Hamiltonian of the form
\begin{align} \label{hamiltonian}
  H = &\sum_{i=1}^N J_i \sigma_z^{(i)} + \sum_{i=1}^N \sum_{j>i}^N J_{ij} \sigma_z^{(i)}\sigma_z^{(j)} \nonumber \\
  + & \sum_{i=1}^N \sum_{j>i}^N\sum_{k>j}^N J_{ijk} \sigma_z^{(i)}\sigma_z^{(j)}\sigma_z^{(k)} + \cdots\,.
\end{align}

The parity transformation maps each $k$-body term in the Hamiltonian to a
physical qubit.  This introduces more degrees of freedom than the original
Hamiltonian, so we need to place some constraints to make sure that the mapped
problem is equivalent to the original.  In previous work \cite{compiler_paper}, some examples were
given of valid parity compilations using plaquettes (i.e., triangles and
squares), where each plaquette corresponds to a constraint.  By picking the
right combination of the possible squares and triangles, the required constraint
space for the parity mapping can be implemented.  In short, the compilation
process ensures that the mapping between the logical and the physical qubits is
sound.

We propose a compilation algorithm that is based on building a rectangular
layout of plaquettes one layer at a time. Each layer consists of a one-dimensional sequence of plaquettes and adds at least one
constraint.  By using only three out of the four possible triangle orientations,
we show how each qubit on the interior can be represented by a product of qubits
on the boundary of the rectangle.  Making use of this boundary mapping, any
constraint can be represented as a combination of boundary qubits and new qubits
that will be placed on the layer exterior to the boundary.  We then show how to
implement that constraint by picking the right combination of squares and triangles
along the exterior of the boundary, possibly using some ancilla qubits.  In many
cases, multiple constraints can be implemented within one layer, and we try to
implement them one after another.  However, if no more constraints can be
implemented in a layer, any remaining space will be filled up with plaquettes to keep the rectangle structure.  Generally, for each layer, there
are several choices when picking the next constraint, and yet, each choice will
ultimately result in a valid parity layout.  We can thus model compilation as an
ancilla minimization problem, where a trade-off between compilation time and
ancilla numbers can be made.

More concretely, here is a description of the algorithm; the details appear later in the paper.
\begin{itemize}
  \item Compute the constraint space $C$, which ensures a valid parity mapping.
  \item Initialize the plaquette space matrix $P$ (describing the constraints that have been added to the layout) and the boundary map matrix
  $B$ to be empty matrices.
  \item While $P$ does not imply all the constraints in $C$:
  \begin{itemize}
    \item Pick a constraint $c$ in $C$ that is not implied by $P$.
    \item If $c$ has more than $3$ qubits that are not in the current layout,
    add them to the layout using ancillas (or pick a different constraint).
    \item Use the boundary map $B$ to get constraint $c'$ in terms of qubits on
    the boundary and up to three new qubits.
    \item Implement $c'$ using several plaquettes, placing new ancilla qubits or
    ancilla plaquettes as necessary.
    \item Add the newly added plaquettes to $P$ and recompute $B$ using these
    plaquettes to express interior qubits in terms of new boundary qubits.
  \end{itemize}
\end{itemize}

\subsection{Description of the algorithm via an example}
Before looking at the algorithm in more detail, let us show the procedure in action using an example.   We consider the following optimization problem:
\begin{align} 
  H = & J_4 \sigma_z^{(4)} + J_{12} \sigma_z^{(1)}\sigma_z^{(2)} + J_{13} \sigma_z^{(1)}\sigma_z^{(3)} \notag \\
  + & J_{14} \sigma_z^{(1)}\sigma_z^{(4)} + J_{23} \sigma_z^{(2)}\sigma_z^{(3)} + J_{34} \sigma_z^{(3)}\sigma_z^{(4)} \notag \\
  + & J_{123} \sigma_z^{(1)}\sigma_z^{(2)}\sigma_z^{(3)}\label{eq:ex_H}\,.
\end{align}
It contains $4$ logical qubits labeled $1$, $2$, $3$, and $4$.   For each
term in the Hamiltonian, we introduce a physical qubit:
$4$, $12$, $13$, etc.   The physical qubits must
satisfy the cyclic constraints in the interaction hypergraph corresponding to
the problem: the interaction hypergraph contains logical qubits as nodes, and
for each interaction, it contains a hyperedge containing the logical qubits in
that interaction.   This implies that each hyperedge corresponds to a physical
qubit.   A cyclic constraint is then a set of hyperedges containing each node an
even number of times.   The set of constraints on the physical
qubits imposed by the compiled layout must then exactly correspond to these
cyclic constraints. For this Hamiltonian, we need three independent constraints; we will show in the constraint algebra section how to calculate this using linear algebra.

We start with the cycle $12$--$23$--$34$--$14$ shown below:
\begin{center}
    \includegraphics[width=0.25\columnwidth]{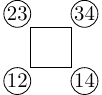}
\end{center}
This was easy because all we needed was a constraint involving three or four
qubits, resulting in a triangle or square.  However, in later stages, we choose
only constraints involving the qubits on the upper-right boundary and up to
three unplaced (i.e., new) qubits.  Thus, we place all the qubits involved in
the new constraint either above the upper boundary or to the right of the right
boundary.  The main idea is that interior qubits such as the physical qubit $12$
can also be included in constraints using the boundary mapping, which maps every
interior qubit to a collection of boundary qubits.  This mapping arises from the
plaquettes that are already placed.  After placing the one constraint as shown
above, the boundary mapping maps the qubit $12$ to $\{23, 34, 14\}$.

For the next constraint we want to place, we could pick $13$--$34$--$14$, which can be added in the following way:
\begin{center}
      \includegraphics[width=0.4\columnwidth]{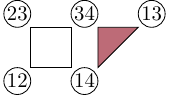}
\end{center}
The plaquettes already have a rectangle structure, so we do not need to add any ancilla
plaquettes.   Now since the qubit $14$ is not on the boundary anymore, we update
our boundary map using the newly added plaquette: $14$ gets mapped to
$\{34, 13\}$, and $12$, which was mapped to $\{23, 34, 14\}$, now gets mapped to
$\{23, 13\}$, by substituting the mapping of $14$.   Now there is only one more
constraint to add, and we pick $4$--$12$--$34$--$123$.   Here, $4$ and $123$ are
new qubits, $34$ is on the boundary, but $12$ is in the interior, so using its
boundary map, we get a new constraint in terms of only boundary and new qubits,
which is: $4$--$23$--$13$--$34$--$123$.   To implement this constraint using plaquettes (see plaquette decomposition of constraints section), we use
an ancilla qubit $x$ as follows:
\begin{center}
    \includegraphics[width=0.4\columnwidth]{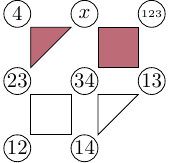}
\end{center}
By combining the plaquettes, the ancilla qubit
cancels out and we get the desired constraint (see constraint algebra section). This finishes the compilation
process for our example.

\section{Detailed algorithm description}
In the rest of the paper, we go over the different steps and features of the
algorithm in more detail.  In the next section, the linear algebra required for
the parity mapping, the boundary map and the plaquette space are discussed.  Then we
demonstrate how constraints consisting of any number of boundary qubits and some
new qubits can be decomposed into plaquettes.  We mention how to handle
constraints with too many new qubits to place directly.  We then discuss how to
optimize the ancilla usage of the algorithm and show some examples of optimized
layouts.  Finally, we show how constrained optimization problems can be compiled
using the proposed algorithm.

\subsection{Parity constraint algebra}
\label{sec:correctness}
\begin{figure}
    \centering
    \includegraphics[width=0.6\columnwidth]{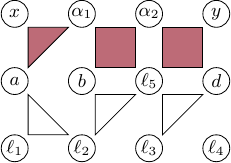}
    \caption{Example showing how the dimension of the constraint space
      implied by the layout increases by one after placing each constraint.
      Suppose that the lower plaquettes are already placed, so our matrix $P$ at
      this time consists of the following row vectors (we can identify bit
      vectors with sets) $\{a, \ell_1, \ell_2\}$, $\{b, \ell_2, \ell_5\}$, and
      $\{d, \ell_3, \ell_5\}$.  The boundary map matrix $B$ is given by the
      following rows: $\{\ell_1, a, b, \ell_5\}$, $\{\ell_2, b, \ell_5\}$, and
      $\{\ell_3, d, \ell_5\}$.  Each row of $B$ is in the rowspace of $P$.
      Suppose that we choose
      $c = \{\ell_1, \ell_3, x, y\}$ as a new constraint to implement.  Since $\ell_1$ and $\ell_3$ are not on
      the boundary, we add corresponding rows from $B$, viz.,
      $\{\ell_1, a, b, \ell_5\}$ and $\{\ell_3, d, \ell_5\}$ to $c$ to get
      $c' = \{a, b, d, x, y\}$, which is implemented as shown in the figure
      above.  If you add all the colored plaquettes, you end up with $c'$ (and
      if you add $\{\ell_1, a, b, \ell_5\}$ and $\{\ell_3, d, \ell_5\}$ to $c'$,
      you get back $c$, which is the constraint we wanted to implement).  Notice
      that even though we added three plaquettes, we also added two ancillas
      $\alpha_1$ and $\alpha_2$, which means effectively that the dimension of
      implied constraint space increased by one, and we are one step closer to
      completing the layout.}
    \label{fig:invariant_ex}
\end{figure}

With the help of linear algebra, we discuss in this section why the algorithm
always finds a correct parity mapping, building on the results from Ref.~\cite{compiler_paper}. All linear algebra discussed here should be understood modulo 2.  For a concrete example complementing
this discussion, see Figure~\ref{fig:invariant_ex}.  We denote by $R(A)$ the
rowspace of a matrix $A$.

We first define a matrix $\cC$, with entries in $\{0, 1\}$. The matrix $\cC$ has
one column for each logical qubit (denoted by numbers) and one column for each parity qubit (denoted by letters).  For
each interaction, $\cC$ has one row with $1$s in the columns corresponding to
logical qubits in that interaction, and one $1$ in the corresponding
parity-qubit column.  This matrix tells us which logical qubits each parity qubit represents.  Here is the matrix for the Hamiltonian we have already seen
before in Eq.~\eqref{eq:ex_H}.
\begin{equation*}
	\begin{array}{cccc|ccccccc}
		1 & 2 & 3 & 4 & a & b & c & d & e & f & g \\
		\hline
		0 & 0 & 0 & 1 & 1 & 0 & 0 & 0 & 0 & 0 & 0 \\
		1 & 1 & 0 & 0 & 0 & 1 & 0 & 0 & 0 & 0 & 0 \\
		1 & 0 & 1 & 0 & 0 & 0 & 1 & 0 & 0 & 0 & 0 \\
		1 & 0 & 0 & 1 & 0 & 0 & 0 & 1 & 0 & 0 & 0 \\
		0 & 1 & 1 & 0 & 0 & 0 & 0 & 0 & 1 & 0 & 0 \\
		0 & 0 & 1 & 1 & 0 & 0 & 0 & 0 & 0 & 1 & 0 \\
		1 & 1 & 1 & 0 & 0 & 0 & 0 & 0 & 0 & 0 & 1
	\end{array}
\end{equation*}
The first row of this matrix represents the equation $a = 1$, the second row $b = 1 + 2$, the third row $c = 1 + 3$, and so on ($1, 2, 3$ are logical-qubit labels, not numbers). We get these equations by multiplying the matrix by the column vector $(1, 2, 3, 4, a, b, c, d, e, f, g)^T$ on the right and equating the product to the zero vector.  This actually gives us equations $1 + a = 0$, $1 + 2 + b = 0$, etc., but since the algebra is modulo $2$, we can add $a, b, \ldots$, respectively, to both sides of these equations to get the equations in the form $a = 1$, $b = 1 + 2$, and so on.
Once we perform Gaussian elimination on this matrix, we obtain the cyclic
conditions that our layout must enforce.  For example, the above matrix is
transformed into the following matrix after Gaussian elimination:
\begin{equation*}
	\begin{array}{cccc|ccccccc}
		1 & 2 & 3 & 4 & a & b & c & d & e & f & g \\
		\hline
		1 & 1 & 0 & 0 & 0 & 1 & 0 & 0 & 0 & 0 & 0 \\
		0 & 1 & 1 & 0 & 0 & 1 & 1 & 0 & 0 & 0 & 0 \\
		0 & 0 & 1 & 1 & 0 & 0 & 1 & 1 & 0 & 0 & 0 \\
		0 & 0 & 0 & 1 & 0 & 1 & 1 & 1 & 0 & 0 & 1 \\
		\hline
		0 & 0 & 0 & 0 & 1 & 1 & 1 & 1 & 0 & 0 & 1 \\
		0 & 0 & 0 & 0 & 0 & 1 & 1 & 0 & 1 & 0 & 0 \\
		0 & 0 & 0 & 0 & 0 & 0 & 1 & 1 & 0 & 1 & 0 \\
	\end{array}
\end{equation*}
The lower part of this matrix denotes all the rows containing zeros in the logical-qubit positions.  And the lower-right part gives us the constraints on parity qubits that our layout must satisfy.  E.g., the last row gives us $c + d + f = 0$; we need to make sure that the plaquette constraints on parity qubits in our layout imply this constraint. Let us denote the lower-right submatrix by $C$ (for constraints).

During the execution of the algorithm, we maintain a matrix-variable $P$ that denotes the
plaquette space of the currently placed plaquettes.  The matrix variable $P$
contains one column for each qubit that is placed so far, which means that some
columns of $P$ correspond to parity qubits and the others correspond to ancilla
qubits.  

All ancilla columns are on the left of $P$.  Note that the algorithm description implicitly says that every time you add a qubit, a column gets added to the matrix variable $P$; if this qubit is an ancilla, a column gets added to the left, otherwise to the right, so that all ancilla columns are on the left of $P$.

Every plaquette constraint the algorithm adds to the layout appears as one row in the matrix-variable $P$.  Since one constraint $c$ in $C$ that is considered in each iteration will result in several plaquette constraints getting added to the layout, several rows get added to the matrix variable $P$ in each iteration.

Denote by $P_f$ the value of $P$ at the end of the algorithm.  After performing
Gaussian elimination on $P_f$, we get the following row-echelon matrix, which we denote by
$P_f^{(G)}$:
\begin{equation*}
  \begin{array}{c|c}
    \text{ancilla} & \text{parity}\\
    \hline
    \mathbf{A_f} & \mathbf{A'_f} \\
    \hline
    \mathbf{0} & \mathbf{C'_f}
  \end{array}
\end{equation*}
This means that the constraints on parity qubits enforced by the plaquettes in
the output of the algorithm are given by $R(C'_f)$.  
The condition in the loop statement ``While $P$ does not imply all the constraints in $C$'' of our algorithm essentially performs this Gaussian elimination and checks if $R(C'_f)$ includes $R(C)$.
We
want to show that $R(C'_f) = R(C)$.  
In words, we
want that after eliminating all the ancillas, our plaquettes give us exactly the
constraints given by $C$.

We make an important basic observation related to Gaussian elimination here.
Suppose we denote by $R_f^{(P)}$ the set of all row vectors containing zeros at
the ancilla positions.  Then, irrespective of how we do the Gaussian elimination
above to get $P_f^{(G)}$, the rowspace of the lower submatrix of $P_f^{(G)}$ is
equal to $R_f^{(P)} \cap R(P_f)$.

Along with $P$, we also maintain a matrix $B$ corresponding to the boundary map
that has one row for each interior qubit $q$: that row contains a $1$ in column
$q$ and the only other $1$'s that appear are in the columns corresponding to
boundary qubits (which can be ancillas).  The crucial property of $B$ is that each row of
$B$ is in $R(P)$, i.e., each mapping is implied by the plaquettes already
placed.

With this setup, we can see by the following inductive argument
that the rowspaces of $C'_f$ and $C$ are the same.  Since we pick a constraint
$c$ to implement such that $c$ is in the rowspace of $C$ but not in that of $P$, the dimension
of the constraint space implied by $P$ increases by at least one.  Every time we
add a constraint, the number of new ancillas we add is equal to one less than
the number of plaquettes we add.  This ensures that we are not losing any extra
degrees of freedom with respect to parity qubits, i.e., the dimension of the
constraint space implied by $P$ increases by at most one.  Thus we finish the
argument by observing that the dimension of the implied constraint space
increases by exactly one.

Some clarification regarding how $c$ is implemented is in order.  Since $c$
can have qubits that are not on the current boundary of the layout, we use the
boundary map to represent $c$ in terms of boundary qubits; say it gives us $c'$.
More concretely, we add to $c$ the rows of $B$ corresponding to the interior
qubits in $c$, which cancels out the interior qubits leaving only boundary
qubits.  We then implement $c'$ by adding plaquettes to the layout whose vectors
sum to $c'$ (as described in the next section).  See
Fig.~\ref{fig:invariant_ex}.

\subsection{Plaquette decomposition of constraints}
In this section, we show how to place qubits and plaquettes in the new layer
so that, together, the plaquettes imply the constraints on the parity
qubits.  A helpful view to adopt is imagining plaquettes being placed starting
from the upper-left corner in the clockwise direction one-by-one towards the lower-right
corner.  Each time we add a plaquette, the qubits that need to be included in the
constraint are ``transmitted'' through the ancilla qubits. We will also need fixed ancilla qubits, which means that they have a single-body constraint fixing them to the ground state. As we see in
the next paragraph, all the information we need to place a constraint is already
on the boundary.  Then placing a constraint becomes a purely geometric process,
where at each step, the shape of the next plaquette can be determined using
simple local conditions.  We see a few examples demonstrating this process.

By using only three out of the four possible triangle orientations, we ensure that all the information stays available on the boundary.  In this paper, we always expand the rectangle upwards and to the right, which means that we cannot use the triangles shaped like this: 
\begin{center}
      \includegraphics[width=0.2\columnwidth]{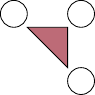}
\end{center}
We call a triangle of this orientation an upper-right triangle, and it is the
only type that blocks information from going to the boundary, because if there
is a qubit in the lower-left corner, it cannot be expressed as a product of
boundary qubits.  We have the other three orientations of triangles, as well as the square at our disposal.

\begin{figure}
    \centering
    \includegraphics[width=0.9\columnwidth]{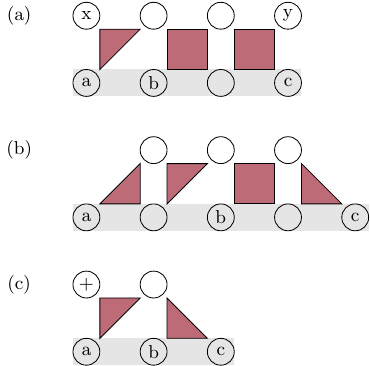}
    \caption{Three examples of constraints consisting of some boundary qubits in
      grey and some new qubits in white.  The unlabeled qubits are not part of
      the constraints as they are in exactly two plaquettes. The new (white) unlabelled qubits are ancillas that can be eliminated using the constraint algebra discussed before (in each figure the number of plaquettes is exactly one larger than the number of white unlabeled qubits). (a) A 5-qubit constraint involving the
      boundary qubits $a$, $b$ and $c$ and two new qubits $x$ and $y$.  (b) A
      3-qubit constraint involving the boundary qubits $a$, $b$ and $c$ that are not directly next to each other.  (c) A
      3-qubit constraint involving the boundary qubits $a$, $b$ and $c$ such that the first two qubits are directly next to each other. To be
      able to draw this constraint a fixed ancilla qubit $+$ is required, which
      is not a degree of freedom.  }
    \label{fig:horizontal}
\end{figure}

A central idea to build a constraint out of plaquettes is that all qubits that are in the constraint appear an odd number of times in the plaquettes and the qubits that are in the constraint appear an odd number of times.  Additionally, the only degrees of
freedom we add are the unplaced qubits in the constraint.  In
Fig.~\ref{fig:horizontal}, some examples of constraints on the horizontal edge
of the rectangle are given, but we discuss a concrete procedure to place such
constraints later in the section.  If the constraint involves any unplaced
qubits, they are added above the first qubit of the constraint and above the
last qubit of the constraint (first and last in the earlier mentioned clockwise
order).  If a constraint involves no new qubits, it may be necessary to use a
fixed ancilla qubit, because an upper-right triangle is not allowed: see
Fig.~\ref{fig:horizontal}~(c).

The geometry arising from this way of placing a constraint has some
properties, for example: if a qubit on the upper boundary is included in the
constraint and there is a plaquette immediately to its left, then that plaquette
must have the following shape (which we call the lower-right triangle):
\begin{center}
      \includegraphics[width=0.2\columnwidth]{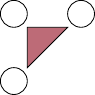}
\end{center}
Whereas, if an upper-boundary qubit $q$ is not included in the constraint and
has a plaquette to its left, then that plaquette must also include $q$ because
the plaquette to $q$'s right cannot exclude $q$ (such a plaquette would be the
forbidden upper-right triangle).  Similar rules apply to the qubits on the right
edge as well.

\begin{figure}
    \centering
    \includegraphics[width=0.7\columnwidth]{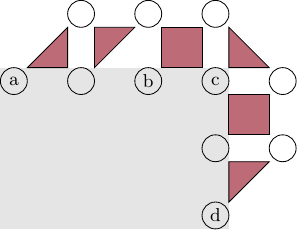}
    \caption{Example of a 4-qubit constraint involving the qubits $a$, $b$, $c$
      and $d$.  In this case the corner qubit $c$ is in three plaquettes, so it
      will be part of the constraint, whereas the unlabeled qubits are in
      exactly two constraints.  If the corner qubit $c$ was not part of the
      constraint, one of the two squares that $c$ is in would have to be a
      triangle, leaving $c$ in only two plaquettes.  In this fashion, a
      constraint with up to three unplaced qubits can be added, one qubit above
      $a$, one qubit diagonally outward from $c$ and one qubit to the right of
      $d$, by turning the respective triangles into squares.  }
    \label{fig:around_corner}
\end{figure}

It is also possible to lay out constraints that involve qubits on both the upper
side as well as the right side of the layout, where the constraint turns around
the corner.  In this case it is possible to add up to three new qubits; an example is given in Fig.~\ref{fig:around_corner}.

We make these ideas concrete by giving an algorithm to place plaquettes for a
given constraint $c$.  Although this algorithm does not handle all cases for the sake of brevity, the ideas
are easily extrapolated to uncovered cases, and we do explain all the uncovered cases before the end of the section.

We first discuss the case when $c$ contains existing qubits that are on the upper boundary and has at most two new qubits.  The uncovered case when they lie on the right boundary is symmetrical.

Denote by $\ell$ the leftmost boundary qubit and by $r$ the rightmost boundary
qubit that is part of $c$.  Place the first new qubit on top of $\ell$ and the
second new qubit on top of $r$.  If spaces above $\ell$ or $r$ are still empty,
add fixed ancilla qubits (this is not necessary in all the cases, but it simplifies the
description of the algorithm).  Denote by $n(q)$ the boundary qubit to the right
of $q$.
\begin{itemize}
  \item Set $q = \ell$.
  \item Repeat the following until $n(q) = r$:
  \begin{itemize}
    \item If $n(q)$ is not in $c$, add a square plaquette.
    \item Else add a lower-right triangle plaquette.
    \item Set $q = n(q)$.
  \end{itemize}
    \item Add a final square plaquette
\end{itemize}
The plaquette that is added is always the plaquette above $q$ and $n(q)$.

There are some final uncovered cases. If $c$ contains qubits from both the upper and the right boundary, we essentially perform the same steps as above
on the separate boundaries, and handle the corner afterwards.  This gives rise to several other cases depending on whether the corner qubit is part of the constraint $c$ or not, but as we mentioned earlier, the same ideas extend.  In the end, the idea is to propagate the information towards the corner by adding plaquettes one-by-one and enforce the constraint on desired qubits by a final plaquette.

\subsection{Addition of new qubits}
If a constraint has too many unplaced qubits, it cannot be directly
implemented, as the number of degrees of freedom that a layer can add is
limited.  This can be remedied either by adding some other constraint first,
which places some of the unplaced qubits, or by adding an artificial constraint
to the layer for the sole purpose of adding a new qubit.  This can, for example,
be done by placing a new qubit above the upper left corner of the layout and
filling the whole horizontal edge with squares.  This effectively adds one
degree of freedom, because it adds exactly one more qubit than plaquettes.  The
filling with squares is important to make sure that all information is still
accessible on the new boundary of the rectangle. 

\begin{figure}
    \centering
    \includegraphics[width=0.25\columnwidth]{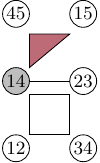}

    \caption{A compilation of a problem with $5$ interactions and $5$ logical qubits, which requires a single parity constraint between all $5$ qubits. The black line shows the separation between the first and second layer. The $5$ white qubits corresponds to Hamiltonian interactions (e.g., $12$ corresponds to the interaction $\sigma_z^{(1)}\sigma_z^{(2)}$) and the grey qubit to an ancilla. The white constraint is placed first and the ancilla is necessary to break up the long constraint into shorter constraints. The red constraint then finishes the layout. }
    \label{fig:ex_ancilla}
\end{figure}

This process of adding qubits by adding artificial constraints is especially important to compile problems that have long constraints that cannot be rewritten as short constraints. For example, imagine a constraint matrix $C$ that has a single row, with five non-zero parity qubit entries, this example is shown in Fig.~\ref{fig:ex_ancilla}. This constraint can never be represented by a single square or triangle, because a plaquette consists of a maximum of four parity qubits. The only solution is to start with an artificial constraint that introduces an ancilla that breaks down the long constraint into two shorter constraints and then implement the constraint in the next layer.

\subsection{Ancilla minimization}
\begin{figure}
    \centering
    \includegraphics[width=\columnwidth]{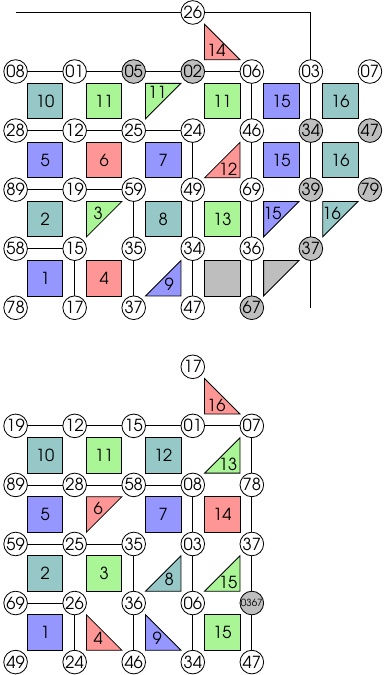}

    \caption{Two different compilations of a problem represented by a
      Hamiltonian with $25$ interactions and $10$ logical qubits.  The black
      grid shows the different layers that build up the rectangle. The $25$
      white qubits corresponds to Hamiltonian interactions (e.g., $78$
      corresponds to the interaction $\sigma_z^{(7)}\sigma_z^{(8)}$) and the
      grey qubits to ancillas.  The plaquette labels number the $16$ required
      constraints, where some constraints require more than one plaquette and
      neighboring constraints are plotted in different colors.  In the top figure, there are also
      two grey plaquettes that are not directly helping implement a constraint, but
      are necessary for bringing qubit information to the boundary.  Unnecessary
      grey plaquettes are removed from the outer layers of the layout.  }
    \label{fig:ex_compilation}
\end{figure}

In the previous sections, we have already highlighted that there are several choices in the algorithm that all lead to different valid parity layouts. For example, the choice of the next constraint to add to the layout and how to place the new qubits in that constraint. In practice, the goal of compilation is to find a layout that is as small as
possible, while still correctly mapping the problem. In an ideal case, each
required constraint can be represented by a single plaquette and all those
plaquettes can be connected to form a complete layout.  However, in many cases
such a layout may not exist, or may be computationally hard to find. This means that there is a trade-off between solution quality (in terms of number of
ancillas) and computation time.  In Fig.~\ref{fig:ex_compilation}, two different
ways to compile the same problem are shown, using different numbers of ancillas.
The more optimized version requires only a single ancilla qubit, because only
one constraint requires two plaquettes and all other constraints are implemented
by a single plaquette.

By iterating over the different options it is often possible to fill the first
layers in a dense way (few plaquettes per constraint).  The final constraints
are the most costly, as by that point the required information is spread out
over a large surface.

For the specific problem shown in Fig.~\ref{fig:ex_compilation}, there are $12$ possible three-qubit constraints and $44$ possible four-qubit constraints, which can be found by enumerating all possible cycles. A triangle can have three different orientations (see the section about plaquette decomposition of constraints) and for each orientation we have six different ways to label the qubits. This means that there are $3 \times 6 = 18$ different ways a three-qubit constraint can result in a triangle. For a four-qubit constraint, there are $24$ different ways to form a square by labeling the qubits. One can start the algorithm by picking any of these triangles and squares as the starting plaquette and all these choices will lead to different final  layouts. This means that there are $12 \times 18 + 44 \times 24 = 1272$ different choices for the starting plaquette. During the rest of the algorithm, there are also choices to be made for which constraint to pick next and where to place the new qubits in the constraint. However, these choices are more restricted due to the condition that the new constraints have to connect to the existing layout. For the example layouts shown in Fig.~\ref{fig:ex_compilation}, thousands of possible final layouts were searched over. Some of these final layouts have many ancillas; in the worst case, each of the $16$ required constraints takes a full layer. This would result in a layout of $16$ by $16$ plaquettes, or $17 \times 17 = 289$ qubits. Without exhaustively going through all options, it is not easy to decide which choices are going to lead to the lowest ancilla numbers. Part of our ongoing research is to find heuristics that optimize this search process and new ways to improve the performance of the algorithm when moving to larger problem instances.

\subsection{Constrained optimization problems}
\begin{figure}
    \centering
        \includegraphics[width=\columnwidth]{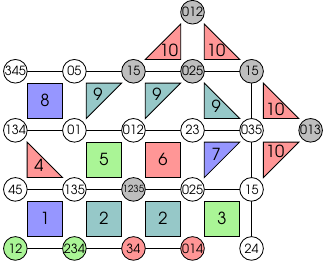}

    \caption{Compilation of a constrained problem represented by a Hamiltonian
      with $16$ interactions and $6$ logical qubits.  The colored qubits
      implement the two side constraints, e.g.  the green qubits implement
      $\sigma_z^{(1)} \sigma_z^{(2)} + \sigma_z^{(2)} \sigma_z^{(3)}
      \sigma_z^{(4)} =a$, where $a \in \{-2, 0, -2\}$.  The black grid shows the
      different layers that build up the rectangle, starting from the four
      qubits in side condition on the bottom row.  The white qubits correspond
      to Hamiltonian interactions that are not in side conditions and the grey
      qubits are ancillas.  The plaquette labels number the $10$ required
      constraints, where some constraints require more than one plaquette and
      neighboring constraints are plotted in different colors.  Unnecessary grey
      plaquettes are removed from the outer layers of the layout.  }
    \label{fig:ex_sum_constraints}
\end{figure}

Constrained optimization problems include side conditions that have to be
satisfied when minimizing the cost function.  This is done by introducing the side conditions as constraints of the form 
\begin{align}\label{side_conditions}
& \sum_{i=1}^N c_i \sigma_z^{(i)} + \sum_{i=1}^N \sum_{j>i}^N c_{ij} \sigma_z^{(i)} \sigma_z^{(j)} \\ \nonumber 
&+ \sum_{i=1}^N \sum_{j>i}^N \sum_{k>j}^N c_{ijk} \sigma_z^{(i)} \sigma_z^{(j)} \sigma_z^{(k)} + \cdots = C_V,
\end{align}
where $C_V$ is the total value of the constraint, in addition to the Hamiltonian of Eq.~\eqref{hamiltonian}.  Product constraints of the form
$\prod_{i} \hat\sigma_z^{(i)} = \pm 1$ can be added in the parity mapping
directly, as shown in Ref.~\cite{compiler_paper}.  General constraints of
polynomial form require a different approach.  Instead of adding these
constraints as penalty terms in the cost function, it is also possible to
satisfy them directly by restricting the driver dynamics.  In
Ref.~\cite{constraint_paper}, this is investigated in detail for the parity
architecture.  The idea is that the initial state of the quantum algorithm
satisfies the constraints and the dynamics do not allow it to evolve into a
state that does not satisfy them, for example, by an exchange Hamiltonian
\begin{equation}
    \label{eq:exch_hamiltonian}
    H_{\textrm{exch.}} = \sum_{\langle i,j \rangle} \tilde{\sigma}^{(i)}_+ \tilde{\sigma}^{(j)}_-  +  h.c.,
\end{equation}
where $\langle i, j \rangle$ are qubits that feature in polynomial constraints.
In order to realise these interactions in quantum machines~\cite{,
  exchange_supercond_martines, exchange_transmon, exchange_rydberg,
  exchange_rydbergZoller}, it is important that they are geometrically local.  This puts the
additional restriction on compilation that the qubits in each side conditions
have to be next to each other.  For our algorithm, one way to do this is by
arranging all the qubits that are in side conditions in groups on the bottom
qubit row of the layout.  In Fig.~\ref{fig:ex_sum_constraints}, an example with
two side conditions, each consisting of two interactions, is compiled.  The first
step of the algorithm puts the four resulting qubits on the bottom row of the
layout and then the algorithm continues exactly as described in the previous
sections.

The drawback of starting from a configuration with all the qubits in side
conditions already placed is that completing the layout may take more layers, as
there is less flexibility in qubit placement.  An alternative approach would be
to initially only place the qubits in a single condition and then try to place
the qubits in other conditions later.  However, at that point it can be
challenging to place the qubits next to each other, or doing so may require
further ancillas.

\section{Conclusion}
Parity compilation using plaquettes can result in advantages in terms of
parallelizability, circuit depth and number of gates.  It also enables
applications in both analog and digital quantum computing.  One major challenge
in exploiting these benefits is to successfully compile to a plaquette layout.
In this work we showed a construction by which higher-order optimization
problems are guaranteed to be mappable to a plaquette layout.  By keeping all qubit information
accessible on the boundary and adding new constraints in each layer, this always
results in a correctly compiled layout.  Near-term quantum devices have limited resources, so optimizing the ancilla usage is crucial.  There are many choices
for the basis of the constraint space as well as the order of the constraints,
but smart search algorithms and heuristics can improve the results in run-time
and ancilla count.  The presented algorithm can bring us one step closer to
realising the potential of various quantum platforms using the parity
architecture.

\section{Acknowledgements}
This study was supported by the Austrian Research Promotion Agency (FFG Project No. 884444, QFTE 2020 and FFG Project No. FO999909249, FFG Basisprogramm). Work was supported by the Austrian Science Fund (FWF) through a START grant under Project No. Y1067-N27 and the SFB BeyondC Project No. F7108-N38. For the purpose of open access, the author has applied a CC BY public copyright licence to any Author Accepted Manuscript version arising from this submission. This project was funded within the QuantERA II Programme that has received funding from the European Union's Horizon 2020 research and innovation programme under Grant Agreement No. 101017733.

%

\end{document}